\documentclass[useAMS,usenatbib]{mn2e}           
\input{psfig.sty}                                

\begin{document}
\title{Nonradial pulsation of the $\delta$ Scuti star UV Trianguli }

\author[Ai-Ying Zhou et al.]
       {Ai-Ying Zhou,$^1$ Zong-Li Liu$^1$ and E. Rodr\'{\i}guez$^2$
       \\
      $^1$National Astronomical Observatories, Chinese Academy of
Sciences,
      Beijing 100012, China, E-mail: aiying@bao.ac.cn\\
      $^2$Instituto de Astrof\'{\i}sica de Andaluc\'{\i}a, CSIC,
      P.O. Box 3004, E-18080 Granada, Spain, E-mail: eloy@iaa.es}

\date{Accepted ~~~~~~~~~~~~. Received 2002 February 00; in original form ~~~~~~~~ }

\pagerange{\pageref{firstpage}--\pageref{lastpage}}
\pubyear{2002}

\maketitle
\label{firstpage}

\begin{abstract}
We present the results of a three-year photometric study of
the $\delta$~Scuti star UV Trianguli.
Our data sets consist of 9378 differential measurements in
Johnson $V$ together with a few data collected into
the Str\"{o}mgren {\em uvby$\beta$} system.
UV Tri is at least a biperiodic variable.
The two best-fitting frequencies, $f_1$=9.3298 d$^{-1}$ and $f_2$=10.8513 d$^{-1}$,
are still not the complete set of pulsation frequencies representing
the star's light variations.
A suspected third frequency might present in the star.
Several ``anomalous cycles'' are observed in the light curves.
They seem real, but are aperiodic.
We derive the colour indices and physical parameters for the variable
and conclude that it is a Population~I $\delta$~Sct star with
normal metal abundance ([Me/H]=0.0$\pm$0.1\,dex) evolving on
its main sequence stage at an early evolutionary phase before the turn-off point.
Finally, we compare the observed oscillation frequencies with theoretical models.
The two pulsation modes of UV Tri are likely nonradial to be gravity modes.

\end{abstract}


\section{Introduction}
The detection of multiple modes along with observational
mode identifications raises the real possibility of performing rigorous
asteroseismology on stars other than the Sun.
Stellar oscillation frequencies are sensitive to
stellar structure and the measurement of a full-set of frequencies
should allow an understanding of the internal structure of a star.
This is the way for us to gain insight into the internal stellar physics.
$\delta$ Scuti stars are regularly pulsating variables situated in
the lower classical Cepheid instability strip on or near the main sequence (MS).
In general, the period range lies between 0.02 and 0.25\,d and the spectral types
range from A2 to F2.
The majority of $\delta$ Sct stars pulsate with a number of
nonradial $p$-modes simultaneously excited to low amplitudes,
but some are (pure) radial pulsators with larger amplitudes and
others pulsate in a mixture of radial and nonradial modes.
The $\delta$ Sct stars have been proven to be a class of promising pulsators
for asteroseismology (e.g. Templeton et al.\ 1997;
Pamyatnykh et al.\ 1998; Breger et al.\ 1999; Michel et al.\ 1999;
Templeton, Bradley \& Guzik 2000).
Therefore, it is necessary to detect as many oscillation
frequencies as possible for individual stars.
For this reason, the star UV Trianguli was selected as one of the targets of
a mission dedicated to the investigations of poorly-studied $\delta$ Sct stars.

UV Tri (=GSC 02293-01382, $\alpha_{2000} = 01^{\rm h}32^{\rm m}00^{\rm s}.11$,
$\delta_{2000} = +30^{\circ}21'56^{\prime\prime}.7$, $V$=11.2\,mag, A3;
Rodr\'{\i}guez, L\'{o}pez-Gonz\'{a}lez \& L\'{o}pez de Coca 2000)
was discovered to be a $\delta$ Sct star by Shaw et al.~\cite{shaw} when
they observed its neighbour eclipsing binary V Tri.
They found a period of $\sim$0.1\,d and pointed out the
possibility of multiple periods, but they could not give further
results due to the deficiency of data. Hence UV Tri was observed
from 1999 to 2001. We collected a number of CCD and photoelectric
photometric data in the Johnson $V$ band, as well as some photoelectric
data in Str\"{o}mgren $uvby\beta$. We are interested in the
behaviour of multiperiodicity and amplitude variability.
A preliminary analysis based on part of the data showed that UV Tri
is a low-amplitude biperiodic $\delta$ Sct pulsator. Two pulsation
frequencies, $f_1$=9.3299 d$^{-1}$ and $f_2$=10.8483 d$^{-1}$ with
semi-amplitudes of 0.026 and 0.010\,mag, respectively, basically
fit the light variations of UV Tri (Liu, Zhou \& Rodr\'{\i}guez 2001).

In this paper, we perform a complete data analysis with the aims to
explore additional pulsation frequencies and to check possible
amplitude variations in the detected frequencies over the observing years.
Section~2 contains an outline of the observational journal and data reduction.
We derive the astrophysical parameters using the $uvby\beta$ indices
in Section~4.  An attempt is also made in this section to model the
pulsations by calculating an appropriate stellar evolution and
oscillation model.
Our main results are discussed and summarized in Section~5.

\section{Data acquisition}
\label{sect:Obs}
The $V$ and $uvby$ observations of UV Tri were secured at two observatories
between 1999 November 17 and 2001 November 22.
The data consist of 9378 measurements in Johnson $V$ band (165.1\,h) and
55 in Str\"omgren $uvby$ colours, collected in 35 observing nights.
Additionally, a few $\beta$ data were also obtained.
A journal of the observations is given in Table~\ref{Tab:log}.

\subsection{Johnson $V$ photometry}
From 1999 November 17 to 2000 January 14,
Johnson $V$ photometry of UV Tri was obtained with
the light-curve survey CCD photometer (Wei, Chen \& Jiang~1990; Zhou et al.\ 2001)
mounted on the 85-cm Cassegrain telescope at the Xinglong Station of
the Beijing Astronomical Observatory (BAO) of China.
The photometer employed a red-sensitive Thomson TH7882 576$\times$384 CCD
with a whole imaging size of 13.25$\times$8.83~mm$^{2}$
corresponding to a sky field of view of $11'.5\times7'.7$, which allows
sufficient stars to be toggled in a frame as reference.
Depending on the nightly seeing the integration times varied from 20 to 60\,s.
During this observing run, the following reference stars in the field of UV Tri
were simultaneously monitored:
C1=GSC 02293-01028 was used as the main comparison star,
C2=GSC 02293-01422, C3=GSC 02293-01027, C4=GSC 02293-01461,
C5=GSC 02293-01456 and C6=GSC 02293-01021 were used as check stars.
The magnitude differences between the reference stars yielded a typical
accuracy of 0.010 to 0.006\,mag.
Before constructing differential time series for the target variable,
special emphasis was made to ensure the constancy of the comparison stars.
Consequently, GSC 02293-01021 was discovered to be a new
W Ursae Majoris-type variable (Liu et al.\ 2000), while the others were
detected as nonvariables within the observational errors.
Finally the remained five reference stars were selected to produce
differential magnitudes for the variable by taking their mean combination as
UV Tri$-$(C1+C2+C3+C4+C5)/5.
Atmospheric extinction was ignored because of the proximity between the
selected objects.
The procedures of data reduction, including on-line bias subtraction,
dark reduction and flatfield correction, are outlined in Zhou et al.\ \cite{zhou01}.

From 2000 October 24 to 2001 November 22, UV Tri was reobserved
with the three-channel high-speed photoelectric photometer (Jiang \& Hu 1998)
attached to the same telescope. This detector is used in
Whole Earth Telescope campaigns (WET; Nather et al. 1990).
GSC 02293-01331 ($\alpha_{2000} = 01^{\rm h}32^{\rm m}12^{\rm s}.46$,
$\delta_{2000} = +30^{\circ}15'35^{\prime\prime}.8$, $V$=10.71) was
chosen as the comparison star.
The variable, comparison star and sky background were simultaneously monitored in
continuous 10-s intervals through a standard Johnson $V$ filter throughout
the two observing seasons.
The typical observational accuracy with the three-channel photometer is
$\sim$0.005\,mag (Zhou 2001).
Most of the $V$ data have been merged into 60-s bins.

\subsection{Str\"{o}mgren $uvby\beta$ photometry}
In 2000 simultaneous photometric $uvby$ observations of UV Tri were
carried out using the six-channel $uvby\beta$ spectrograph photometer
attached to the 90-cm telescope at the Observatorio de Sierra Nevada (OSN),
Granada, Spain (Rodr\'{\i}guez et al.\ 1997). Additionally, a few data using
the Crawford H$_{\beta}$ system were also collected.
This colour photometry allows us the determination of the corresponding
photometric indices and hence astrophysical parameters for the variable.

During these observations, HD 9483 ($V$=8.1, A3) was used as the
main comparison star with HD 9445 ($V$=8.5, F0) and
HD 8826 ($V$=8.0, F0) as check stars. The spectral types and
colour indices of the three stars locate them inside the instability region of
the $\delta$~Sct and/or $\gamma$ Dor type pulsators.
However, during the observations reported here, no variability was detected
for any of the comparison stars within the observational errors.

To transform our data into the standard system we have used the same procedure
described in Rodr\'{\i}guez et al.\ (1997). In the present work
10 standard stars were observed for purposes of calibration.
The standard deviations obtained in the transformation equations were
0.011, 0.007, 0.009, 0.009 and 0.008\,mag for $V$, $b-y$, $m_1$, $c_1$
and $\beta$, respectively.
The results, together with the standard deviations and the number of points
collected for each object, are listed in Table~\ref{Tab:uvby}.
As can be seen, our results are in good agreement with those available
in the bibliography (Hauck \& Mermilliod 1998).
Due to the small number of $uvby$ data, $y$ points were not merged into the $V$ data.
%
%
\begin{table}
  \caption[]{Journal of Johnson $V$ photoelectric and CCD measurements
  together with Str\"{o}mgren $uvby$ photometry of UV Tri. The JD is in
2450000+ days.}
  \label{Tab:log}
  \begin{center}\begin{tabular}{ccccl}
  \hline
Set & Night(UT)  & JD   & \hspace{-5pt} Time interval(d) & \hspace{-6pt}Points, Filter \\
  \hline
1999   & 1999.11.17  &1500  & 0.1051 & 122, $V$ \\
       & 1999.11.18  &1501  & 0.1714 & 199, $V$ \\
       & 1999.11.24  &1507  & 0.1262 & 133, $V$ \\
2000   & 2000.01.08  &1552  & 0.0709 & ~89, $V$ \\
       & 2000.01.14  &1558  & 0.1477 & ~95, $V$ \\
       & 2000.01.29  &1573  & 0.1210 & ~14, $uvby$ \\
       & 2000.02.02  &1577  & 0.1120 & ~21, $uvby$ \\
       & 2000.02.03  &1578  & 0.1022 & ~20, $uvby$ \\
       & 2000.10.24  &1842  & 0.2948 & 419, $V$ \\
       & 2000.10.25  &1843  & 0.3840 & 554, $V$ \\
       & 2000.10.26  &1844  & 0.0493 & ~72, $V$ \\
       & 2000.10.29  &1847  & 0.3687 & 498, $V$ \\
       & 2000.10.31  &1849  & 0.4006 & 569, $V$ \\
       & 2000.11.01  &1850  & 0.4048 & 584, $V$ \\
       & 2000.11.02  &1851  & 0.2333 & 337, $V$ \\
       & 2000.11.04  &1853  & 0.2682 & 378, $V$ \\
       & 2000.11.05  &1854  & 0.1278 & 123, $V$ \\
       & 2000.11.06  &1855  & 0.0424 & ~62, $V$ \\
       & 2000.11.07  &1856  & 0.0576 & ~74, $V$ \\
       & 2000.11.10  &1859  & 0.1263 & 176, $V$ \\
       & 2000.11.11  &1860  & 0.1882 & 268, $V$ \\
       & 2000.11.12  &1861  & 0.3223 & 465, $V$ \\
       & 2000.11.13  &1862  & 0.3180 & 459, $V$ \\
       & 2000.11.14  &1863  & 0.3569 & 515, $V$ \\
       & 2000.11.17  &1866  & 0.3381 & 435, $V$ \\
       & 2000.11.19  &1868  & 0.1779 & 235, $V$ \\
       & 2000.11.20  &1869  & 0.2842 & 367, $V$ \\
2001   & 2001.10.17  &2200  & 0.1013 & 128, $V$ \\
       & 2001.10.18  &2201  & 0.4138 & 593, $V$ \\
       & 2001.10.22  &2205  & 0.2595 & 367, $V$ \\
       & 2001.11.06  &2220  & 0.0645 & ~94, $V$ \\
       & 2001.11.07  &2221  & 0.1209 & 174, $V$ \\
       & 2001.11.08  &2222  & 0.1409 & 204, $V$ \\
       & 2001.11.09  &2223  & 0.0742 & 108, $V$ \\
       & 2001.11.22  &2236  & 0.3403 & 482, $V$ \\
  \hline
  \end{tabular}
  \end{center}
\end{table}
%
%
\begin{table}
  \caption[]{$uvby\beta$ indices obtained for UV Tri and the comparison
  stars. The pairs below the star names are the measurements collected for
  each object in $uvby$ and $\beta$, respectively. In the bottom part,
  the values listed in the Hauck \& Mermilliod (1998) catalogue
  are shown for comparison.}
  \label{Tab:uvby}
  \begin{center}\begin{tabular}{crrrrr}
  \hline\noalign{\smallskip}
\multicolumn{1}{c}{ Object} &
\multicolumn{1}{c}{ $V$ }   &
 $b-y$ & \multicolumn{1}{c}{$m_1$ }  &\multicolumn{1}{c}{ $c_1$} &
\multicolumn{1}{c}{$\beta$}   \\
  \hline\noalign{\smallskip}
UV Tri    & 11.085 & 0.215 & 0.169 & 0.783 & 2.775 \\
(55,~4)   &   .021 &  .015 &  .014 &  .041 &  .016 \\
HD 9483   &  8.091 & 0.116 & 0.161 & 0.985 & 2.835 \\
(91,~7)   &      4 &  .004 &  .005 &  .015 &  .009 \\
HD 9445   &  8.482 & 0.252 & 0.162 & 0.779 & 2.722 \\
(20,~4)   &      4 &  .005 &  .005 &  .015 &  .007 \\
HD 8826   &  7.998 & 0.274 & 0.148 & 0.459 & 2.669 \\
(21,~4)   &      5 &  .004 &  .004 &  .011 &  .012 \smallskip\\
HD 9483   &  8.06  & 0.120 & 0.141 & 0.979 & 2.838 \\
HD 9445   &  8.50  & 0.255 & 0.167 & 0.788 & 2.717 \\
HD 8826   &  8.01  & 0.267 & 0.163 & 0.456 & 2.671 \\
  \noalign{\smallskip}\hline
  \end{tabular}
  \end{center}
\end{table}

\section{Data analysis}
\label{sect:FA}

\subsection{Frequency solution}
To find a complete set of pulsation frequencies of UV Tri,
we merged all the data collected from 1999 to 2001.
The frequency analysis was carried out by using the programmes {\sc period98} (Breger 1990;
Sperl 1998) and {\sc mfa} (Hao 1991; Liu 1995), where
single-frequency Fourier transforms and multifrequency least-squares fits
were processed.
The two programmes use the Discrete Fourier Transform method (Deeming 1975)
and basically led to identical results.
To judge whether or not a peak is significant in the amplitude spectra we
followed the empirical criterion of Breger et al.\ (1993),
that an amplitude signal-to-noise (S/N) ratio larger than 4.0 usually
corresponds to an intrinsic peak of the variable.
Hence the noise levels at each frequency were computed using the residuals
at the original measurements with all the trial frequencies
prewhitened. Then the confidence levels of the frequencies were
estimated following Scargle (1982).
Note that the S/N criterion assumes a good spectral window typical of
multisite campaigns.
However, in the case of single-site observations, the noise level can be
enhanced by the spectral window patterns of the noise peaks and possible
additional frequencies. Therefore a significant peak's S/N value might be
a little less than 4.0.

First, we need to check the constancy of the comparison star GSC 02293-01331
used in the photoelectric observations.
Unfortunately, no differential photometry was made for it.
We analysed the amplitude spectrum of the comparison star in
the frequency range 0--50 d$^{-1}$.
The data from the 9 photometric nights
on 2000 October 24, 25 and 31; November 1, 4, 12, 13, 14 and 20 were thus used.
A forced Fourier analysis produced four peaks in the lower frequency region:
3.0125, 2.6722, 4.8577 and 1.7974 d$^{-1}$ with amplitudes of
0.064, 0.025, 0.020 and 0.019\,mag, respectively.
We are unable to judge whether these peaks are or not intrinsic to
the comparison star depending on their S/N values alone.
The first (highest) peak at 3.0 d$^{-1}$ is due to the
atmospheric extinction,
which could not be completely corrected in the case of lacking a check star.
The other peaks might be mainly attributed to atmospheric or instrumental effects.
We keep in mind that caution must be taken in detecting frequencies
in the region lower than 4 d$^{-1}$, as pointed out by Breger (1994),
that 0--4 d$^{-1}$ can be regarded as the garbage heap for
observational problems such as incomplete extinction corrections and
equipment changes.
In order to reduce the low-frequency noise, these four frequencies have
been removed from the differential measurements for UV Tri below.

\begin{figure}
   \vspace{2mm}
   \hspace{2mm}\psfig{figure=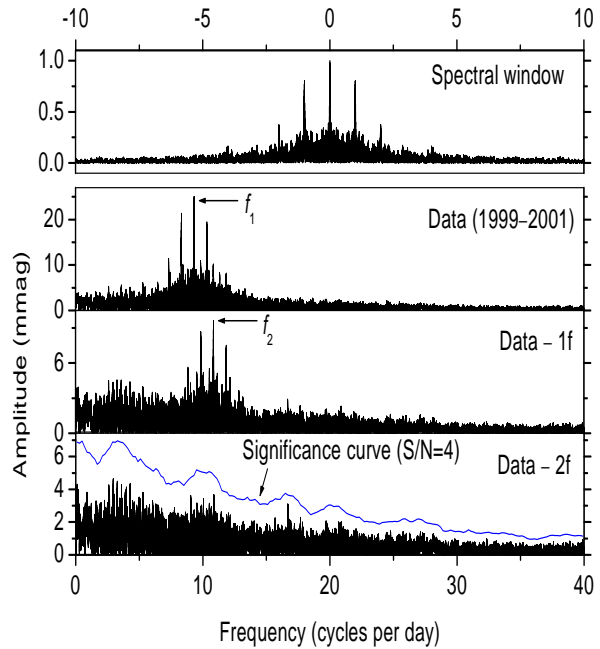,width=90mm,height=95mm,angle=0.0}
   \caption{The spectral window and amplitude spectra of UV Tri (1999--2001).
   Note the contaminated pattern around $\sim$ 4 d$^{-1}$ is visible
   in last two panels, which might be noised by instrumental,
   atmospheric and other effects.  }
   \label{Fig:power}
\end{figure}

\begin{figure*}
   \vspace{2mm}
\hspace{2mm}\psfig{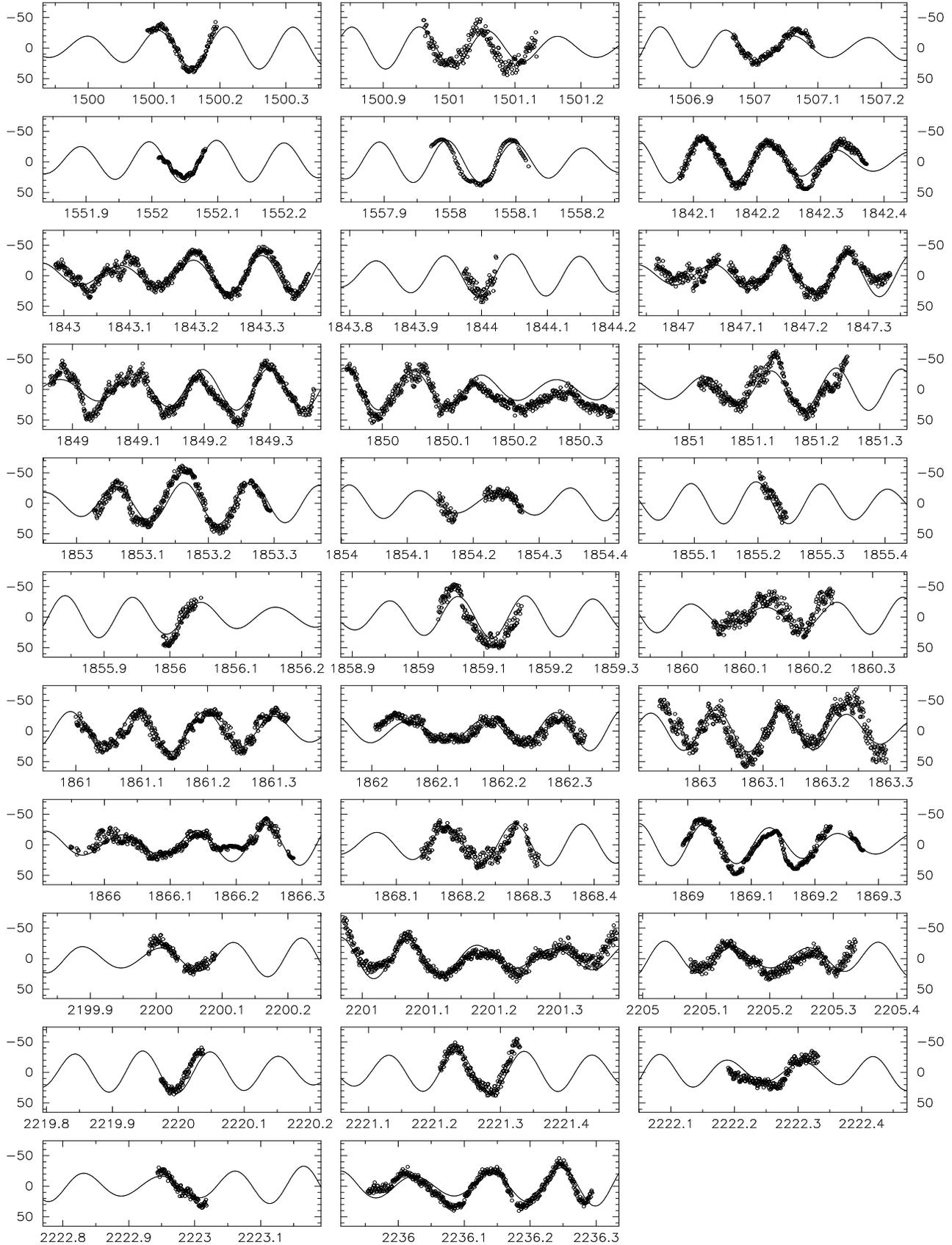}
   \caption{Observed $V$ differential light curves (circles) during 1999--2001
   together with the fitted two frequencies represented as solid lines.
   Abscissa in HJD 2450000+ days, ordinate in mmag.}
   \label{Fig:lightcurve-V}
\end{figure*}

Fourier analyses of all the data for UV Tri show two dominate
frequencies at $f_1$=9.3298 d$^{-1}$ and $f_2$=10.8513 d$^{-1}$,
verifying the previous detection (Liu et al.\ 2001) with an exception
of the term at 3.6 d$^{-1}$.
Subsequent analyses of the residuals after prewhitening the two frequencies
show that the domain in 0--7 d$^{-1}$ was highly contaminated.
The higher noise in the region lower than 4 d$^{-1}$ mostly reflects
the effects caused by instrumental drifts, atmospheric changes (e.g. irregularities
in sky transparency and scintillation noise) and other similar phenomena not caused
by the target star.
For safety, we stopped searching for pulsation in this lower-frequency region.
We have noted the existing aliases in the data by taking into account the integer
multiples of 0.00137, 0.033 and 1.0 d$^{-1}$ corresponding to two-year alias,
monthly alias and daily alias, respectively.
For the judgement of significant peaks the amplitude spectra and spectral window
are shown in Fig.~\ref{Fig:power}, in which each spectrum panel corresponds to
the residuals with all the previous frequencies prewhitened.
The last panel (`Data -- 2f') shows the residuals after subtracting the fit
of the two outstanding frequencies, together with the significance
curve -- a spline-connected points -- four times the noise points obtained in
a spacing of 0.2 d$^{-1}$.
The observational noise is frequency-dependent and it was defined as
the average amplitude of the residuals in a range of 1 d$^{-1}$ (11.57 $\mu$Hz),
close to the frequency under consideration.
As a final result, the pulsation frequencies $f_1$=9.3298 d$^{-1}$ and
$f_2$=10.8513 d$^{-1}$ with
amplitudes of 0.0254 and 0.010\,mag, respectively, are confirmed to be
intrinsic to UV Tri.
So the variable is currently a biperiodic $\delta$ Sct pulsator.
The two frequencies fit the light curves with a standard deviation of
$\sigma$=0.013\,mag.
Fourier parameters of the best-fitting sinusoids are
listed in Table~\ref{Tab:freq}, where the errors in frequencies, amplitudes and
phases were estimated assuming the root-mean-square deviation of the
observational noise to be of 0.013\,mag, the standard deviation of the fit,
through the formulae of Montgomery \& O'Donoghue~(1999). We used the real nights
with data for the time baseline rather than the span of observations,
so that the errors would not be underestimated.
The differential light curves along with the fit are presented
in Fig.~\ref{Fig:lightcurve-V}.

From Figs.~\ref{Fig:power} and~\ref{Fig:lightcurve-V},
the detection of pulsation frequencies seems not complete.
The residual spectrum (`Data -- 2f' panel in Fig.~\ref{Fig:power}) is
still not consistent with white noise.
This indicates that other pulsation frequencies are possibly present in the data.
Poor fits on some nights, such as HJD 2451842, 2451849 and 2451850,
also show this possibility.
The region around 10 d$^{-1}$ might be affected by
potential amplitude modulation of $f_1$ and $f_2$ and by additional
frequencies very close to $f_1$.
We noted a peak at 16.6862 d$^{-1}$ ($f_3$) with an amplitude of
0.0031\,mag (S/N = 3.4). It is probably
a candidate frequency
as its height is obviously higher than its neighbour peaks.
Inspection of the light curves showed several peculiar cycles
on HJD 2451843.1, 2451850.05, 24511851.12, 2451863.2 and 2452201.3,
in which irregular portions of the light curves occurred on the
ascending branches near the maxima.
They were not fitted at all by the two detected frequencies.
Fitting by taking $f_3$ into account did not reproduce these unusual cycles and
did not improve the goodness of fitting remarkably.
We paid close attention to these asymmetric shapes in the light curves
because we found them immediately after the end of observations in those nights.
These unusual cycles seemed not related with non-photometric conditions,
instruments or the comparison star.
They are probably real and aperiodic.
The effect may be referred to a similar phenomenon of ``anomalous cycles'' reported by
Papar\'{o} et al.\ (2000) in their study of the $\delta$ Sct star 57 Tauri.
These authors described the phenomenon as single unusual cycles of light variation
with larger amplitude than the surrounding cycles or with asymmetric shapes.
Hence the amplitudes are larger than expected from multifrequency solution of
the light curve. They also pointed out that
another $\delta$ Sct star XX Pyxidis seems to show anomalous cycles.
However, Handler (2001) argued that the hypothesis that XX Pyx shows
anomalous cycles is quite unlikely. Handler suggested that single anomalous cycles of
$\delta$ Sct stars are physically implausible. The possibility of anomalous cycles
occurring in pulsating stars should not be dismissed on superficial grounds,
but the evidence for it needs to be accumulated and critically scrutinized.
The case of anomalous cycles in UV Tri, which needs further
observational check, supports the suggestion that the phenomenon is intrinsic to
$\delta$ Sct stars. Investigation of possible causes and implications
of the anomalous cycles becomes more interesting.
%

%

%
%
\begin{table}
   \caption[ ]{Results of the frequency analysis of UV Tri.
   Epoch in HJD 2451842.0+ days.}
   \label{Tab:freq}
   \begin{center}\begin{tabular}{crrcrc}
   \hline\noalign{\smallskip}
\multicolumn{2}{c}{ Freq.}
                    &Ampl.  &  Phase, Epoch & S/N & Conf.\\
\multicolumn{2}{c}{( d$^{-1}$)}
                    &(mmag) &  (0--1, days) &     & (\%) \\
   \hline\noalign{\smallskip}
 $f_{1}$&  9.3298 & 25.37 &  0.096, .015 &21.2 & 100  \\
        &   .0001 & ~~.18 &  ~.001, ~~~~ &     &      \\
 $f_{2}$& 10.8513 &  9.91 &  0.524, .018 & 9.3 & 100  \\
        &   .0003 & ~~.18 &  ~.001, ~~~~ &     &      \\
   \hline\noalign{\smallskip}
   \end{tabular}
   \end{center}
\end{table}

\subsection{Frequency and amplitude variability}
It is possible to search the data for amplitude and/or frequency variability
on an annual time-scale for UV Tri.
We divided the data into three subsets according to the observing gaps:
data set `1999' included measurements collected between 1999 November 17
and 2000 January 14 (5 nights), data set `2000' from 2000 October 24 to
November 20 (19 nights), while `2001' for those in 2001 (8 nights).
Individual Fourier analyses were then carried out for each subset.
However, we are aware that the Fourier results highly depend on the data structure,
i.e. the number of data points, sampling rate, time span, etc.
Furthermore, Fourier analysis assumes the frequency and amplitude of a signal
to be constant in the time domain. Therefore, our results for the time-dependence of
frequency and amplitude should be regarded as the averaged values
during the investigated period.
Given the trial frequency values of $f_1$=9.3298 d$^{-1}$ and $f_2$=10.8513 d$^{-1}$,
non-linear least-squares sinusoidal fit to each data set were made.
Our results are presented in Table~\ref{Tab:freq-var},
where the errors were estimated following the methods in previous section.
The standard deviations of fit for the three subsets are
$\sigma$=0.0115, 0.0118 and 0.0108\,mag, respectively.
The fitting quality is generally consistent among the three solutions.
As can be seen, the two frequencies were constant in 1999--2001 within
the calculated errors.
%
\begin{table}
  \caption[]{Fourier parameters of the best-fitting sinusoids for the
three subsets
  of data in 1999--2001.  }
  \label{Tab:freq-var}
  \begin{center}
  \begin{tabular}{cccl}
  \hline\noalign{\smallskip}
Subset&Frequency(d$^{-1}$) &  Amplitude(mmag)  & Phase(0--1)    \\
  \hline
1999  &$f_1$~ ~9.327$\pm$.003 & 28.3$\pm$.7 & .264$\pm$.004 \\
      &$f_2$~ 10.852$\pm$.006 & 13.5$\pm$.7 & .611$\pm$.008 \\
2000  &$f_1$~ ~9.329$\pm$.0002& 26.0$\pm$.2 & .663$\pm$.001 \\
      &$f_2$~ 10.847$\pm$.001 & 10.1$\pm$.2 & .695$\pm$.003 \\
2001  &$f_1$~ ~9.328$\pm$.001 & 24.1$\pm$.4 & .196$\pm$.002 \\
      &$f_2$~ 10.855$\pm$.003 & ~9.0$\pm$.4 & .286$\pm$.007 \\
  \noalign{\smallskip}\hline
  \end{tabular}
  \end{center}
\end{table}

By fixing the two frequencies $f_{1}$=9.3298 d$^{-1}$ and $f_{2}$= 10.8513 d$^{-1}$ and
allowing their amplitudes and phases to change,
we obtained the amplitudes of both frequencies for different years' data.
The details are given in Table~\ref{Tab:ampl-var}.
In the estimated errors range, there is still a difference of the values of
the amplitude of $f_1$ between Tables~\ref{Tab:freq-var} and~\ref{Tab:ampl-var}.
Other values are basically consistent with each other.
One reason causing this difference probably is the shortest data in 1999
because the results in Table~\ref{Tab:freq-var} are the least-squares optimised values
of all three Fourier parameters of the two frequencies.  While the results in
Table~\ref{Tab:ampl-var} are the optimised values of amplitude and phase
when the frequencies are fixed.
Compared with the results in Table~\ref{Tab:freq} obtained for the overall data sets,
which are shown as the solid line in Fig.~\ref{Fig:lightcurve-V},
the amplitudes calculated in Table~\ref{Tab:ampl-var} are
more certain and reasonable than in Table~\ref{Tab:freq-var}.
We did not see anomalous cycles in the light curves of 1999.
Deviations between fits and light curves are relatively smaller in the 1999 data set
than in other two sets. The overall frequency solution is good enough for
the 1999 data set. We believe anomalous cycles did not affect the amplitude in 1999.
The error for amplitudes in 1999 may be underestimated due to shorter data length.
However, even in Table~\ref{Tab:ampl-var}, the amplitude of
$f_2$ in 1999 is obviously higher than in other two years.
In 2000 and 2001, there are many cycles covered by the observations, so we think
that a few disturbances (i.e. the anomalous cycles) among so many
cycles is not enough to affect the amplitude.
Consequently, from Tables~\ref{Tab:freq-var} and~\ref{Tab:ampl-var} we think
that the amplitude of $f_1$ did not change during these three years,
but the amplitude of $f_2$ is larger in 1999 than those in 2000 and 2001.
In view of the higher noise around 10 d$^{-1}$ in the amplitude spectra,
amplitude variability as a cause cannot be ignored and it deserves
further investigation.
%
%
\begin{table}
  \caption[]{Amplitudes (mmag) of the two well-resolved frequencies
  (d$^{-1}$) of UV Tri in 1999--2001.   }
  \label{Tab:ampl-var}
  \begin{center}
  \begin{tabular}{cccc}
  \hline\noalign{\smallskip}
               &    1999     &   2000      &   2001      \\
  \hline
$f_1$= 9.3298 & 25.6$\pm$.7 & 26.1$\pm$.2 & 24.2$\pm$.4 \\
$f_2$=10.8513 & 13.6$\pm$.7 & ~9.7$\pm$.2 & ~9.3$\pm$.4 \\
  \noalign{\smallskip}\hline
  \end{tabular}
  \end{center}
\end{table}

\section{Properties of UV Tri}
\label{sect:phy-par}
In this section, we derive the stellar properties of UV Tri.

\subsection{Physical parameters from $uvby\beta$ indices}
\label{sect:phy-par}
An estimate of stellar physical parameters including absolute magnitude,
surface gravity, effective temperature and other quantities can be
derived by applying suitable calibrations for $uvby\beta$ photometry.
Using the observed colour indices obtained in the present work
(Table~\ref{Tab:uvby}), we can deredden these indices
making use of the dereddening formulae and calibrations for A stars
given by Crawford (1979).
This way, we derive a colour excess of $E(b-y)$=0.048\,mag and the
following intrinsic indices: $(b-y)_0$=0.167, $m_0$=0.185 and
c$_0$=0.774\,mag. Furthermore, deviations from the zero age main sequence
values of $\delta$$m_0$=0.009 and $\delta$c$_0$=0.044\,mag are also found.
These values suggest this variable to be a normal Population I
$\delta$~Sct star with normal abundance in metals and
well-situated into the $\delta$~Sct region in the H-R diagram
as can be seen from fig.\ 8 of Rodr\'{\i}guez \& Breger~(2001).
In fact, a value of the mean metal abundance of [Me/H]=$-$0.014\,dex (or Z=0.019)
can be derived using the calibrations by Smalley~(1993) for metallicity
of A-type stars with $\delta$$m_0$=0.009\,mag.
We note here that the spectral type of UV Tri, A3, the same as that of V Tri,
was estimated by Shaw et al. (1983). No one spectroscopically-determined type
is available.

We further derived the effective temperature, surface gravity,
bolometric magnitude and radius:
$T_{\rm eff}= 7480\pm150$\,K, $\log g= 4.06\pm0.06$,
$M_{\rm bol}=2.46\pm0.20$\,mag ($M_{V}=2.44\pm0.20$\,mag) and
$R$=1.74$\pm$0.23\,R$_{\odot}$.
We used $\beta$, which is free of interstellar extinction effects,
as the independent parameter for measuring temperature (Crawford 1979).
The effective temperature and gravity were determined from
the model-atmosphere calibrations of $uvby\beta$ photometry by
Moon \& Dworetsky (1985).
The bolometric magnitude was found at normal metal abundance.
A bolometric correction, B.C.=0.02\,mag, derived from
Malagnini et al.\ (1986) for $T_{\rm eff}$=7480\,K was assumed.
The radius value was calculated by the radiation law.
For the B.C., Balona (1994) gives 0.05$\pm$0.047\,mag for the temperature,
so the bolometric magnitude of UV Tri becomes:
$M_{\rm bol}$ = 2.49$\pm$0.25\,mag ($\log L/L_{\sun}$ = 0.90$\pm$0.10).
The two values are in agreement within the error range.
Following Crawford \& Mandwewala~(1976) the interstellar reddening is
0.21\,mag for the colour excess of $E(b-y)$=0.048\,mag.
Consequently, a distance modulus D.M.=8.4$\pm$0.2\,mag was obtained.

Moreover, it is possible to gain some insight into the mass and
age of this star using the evolutionary tracks of Claret (1995)
for solar abundances. In this case, values of an evolutionary mass
$M$=1.78$\pm$0.1\,M$_{\odot}$, a bolometric magnitude
$M_{\rm bol}$=2.03$\pm$0.3\,mag and an age of 0.89$\pm$0.05\,Gyr are
found with $T_{\rm eff}$=7480\,K and $\log g$=4.06.
So a mean density of $\bar{\rho}$=0.34$\pm$0.14\,$\rho$$_{\sun}$ was derived.
In addition,
an asteroseismic mass of 1.63\,M$_{\odot}$ can be derived by using the
equation of Arentoft et al.\ (1998) with $M_{\rm bol}$=2.46\,mag.

The results have been checked against two other calibrations using
$uvby\beta$ photometry: firstly, the biparametric calibrations for
stellar mass, radius and surface gravity by Ribas et al.\ (1997) and,
secondly, the absolute magnitude calibration by Domingo \&
Figueras (1999). A good agreement with the above was reached.
In addition, our previously calculated values of $T_{\rm eff}$, $\log g$ and
{\em M} are also supported by the grids of Smalley \& Kupka~(1997)
with [Me/H]=0.0 and by Balona's (1994) calibrations.
Table~\ref{Tab:phy-par} summarizes the parameters obtained for UV Tri.
%
%
\begin{table}
  \begin{center}
  \caption{Dereddened indices and derived physical parameters for
UV Tri. D. M. stands for distance modulus. }
  \label{Tab:phy-par}
  \smallskip
  \begin{tabular}{lclc}
  \hline
Parameter      &  Values (mag)   &   Parameter       &    Values \\
  \hline
$E(b-y)$       & 0.048$\pm$0.01  & age (Gyr)         &  0.89$\pm$0.05 \\
$(b-y)_0$      & 0.167$\pm$0.01  & $M$/M$_{\odot}$   &  1.78$\pm$0.08 \\
$m_0$          & 0.185$\pm$0.01  & $R$/R$_{\odot}$   &  1.74$\pm$0.23 \\
$c_0$          & 0.774$\pm$0.01  & $\log L/L_{\odot}$&  0.90$\pm$0.08 \\
$\delta$$m_0$  & 0.009$\pm$0.01  &$T_{\rm eff}$ (K)  &  7480$\pm$150  \\
$\delta$$c_0$  & 0.044$\pm$0.01  & $\log T_{\rm eff}$&  3.874$\pm$0.009   \\
$M_V$          & 2.44 $\pm$0.2   & $\log g$ (dex)    &  4.06$\pm$0.06     \\
$M_{\rm bol}$  & 2.46 $\pm$0.2   & [Me/H] (dex)      & $-$0.014$\pm$0.1   \\
D. M.          & 8.4~ $\pm$0.2   & $\bar{\rho}/\rho_{\sun}$&0.34$\pm$0.14 \\
  \hline
  \end{tabular}
  \end{center}
\end{table}

\subsection{Possible mode identification}
Here we try to identify the modes for the two detected frequencies in
terms of their pulsation constants and frequency ratio. We also compare them
with existing models.

The physical parameters derived above are used to compute
the observed pulsation constants $Q_{i}$ related to each frequency
by means of the empirical formula of Petersen \& J$\o$rgensen~\cite{pete72}.
We obtained $Q_1$=0.053 and $Q_2$=0.046\,d for $f_1$ and $f_2$.
The $Q$ values derived from $uvby$ photometry have an uncertainty of
$\sim$18 per cent ($\sim$0.01\,d in our case) according to Breger et al.\ (1999).
Taking this and the models of Fitch (1981) into account,
$f_1$ and $f_2$ can be identified as $g_1$ modes
with $l$=2 and 3, respectively.
That is, UV Tri pulsates in low-degree ($l$) nonradial gravity ($g$) modes of
low-order (radial overtone 1).
This result is quite interesting because $\gamma$ Dor-type stars pulsate in
low-degree nonradial gravity modes of high-order and UV Tri locates in the
overlapped region of $\delta$ Sct and $\gamma$ Dor variables (Handler 1999; Guzik et al.\ 2000).
The mode identifications based upon $Q$ values rely on the available models,
of which the Fitch's modes are widely used and are effective.
On the other hand, the results using $Q$ solely might be dramatic (Garrido 2000).

In addition, we can also refer the variable to other
existing models. The observed frequency ratio $f_1/f_2$=0.86
first rules out at least the existence of two radial modes.
According to the empirical period-luminosity-colour relations of
Stellingwerf \cite{stel79} and L\'{o}pez de Coca et al.\ \cite{lope90} and
adopting $M_{\rm bol}$=2.46\,mag and $T_{\rm eff}$=7480\,K,
both $f_1$ and $f_2$ cannot be referred to a radial mode.
Referred to the theoretical frequencies as well as
frequency ratios of the radial oscillation modes of the
2.2\,M$_{\sun}$ model with solar metallicity (Z=0.02) of
Viskum et al.\ \cite{visk98}, our derived mean density
$\bar{\rho}$=0.34$\pm$0.14\,$\rho_{\sun}$ does not suggest a
radial mode at the values of either $f_1$ or $f_2$. Furthermore,
both frequencies are neither present in the 2.0\,M$_{\sun}$
(Z=0.02) model with $l$=0--2 of Templeton et al.\ \cite{temp97}
nor in the models of Breger et al.\ (1999) for the
well-studied $\delta$ Sct star FG Virginis. It is clear that the degree
$l$ for the two modes is at least nonzero.

Consequently, the oscillation nature of both $f_1$ and $f_2$ is
most likely to be purely nonradial.
Unfortunately, current $uvby$ data are too few to
perform an observational mode identification for the variable.
Confirmation for the nonradial nature is needed.

\subsection{A pulsation model}
\begin{figure}
   \vspace{0.2mm}
   \hspace{1mm}\psfig{figure=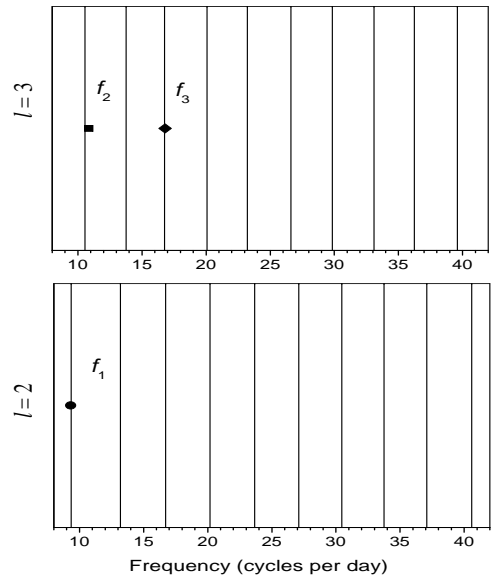,width=85mm,height=90mm,angle=-0.0}
   \caption{Theoretically predicted low-degree ($l$=2, 3) pulsation
   frequencies (vertical lines) of the 1.65\,M$_{\sun}$ model compared with the
   observed frequencies of UV Tri (dot, square and diamond for $f_1$,
   $f_2$ and $f_3$, respectively).}
   \label{Fig:freq-model}
\end{figure}

Here we try to construct an asteroseismic model for UV Tri.
This model tends to identify the star's evolutionary phase and
pulsation modes.
According to the evolutionary status of UV Tri and referring to the
asteroseismic mass calculated above, we computed the theoretical
pulsation frequencies following the minimal asteroseismological
approach (Liu et al.\ 1999; Breger 2000).
A pulsation model, corresponding to the evolutionary stage of
$\log L/L_{\sun}$ = 1.00 and $\log T_{\rm eff}$ = 3.871 ($T_{\rm
eff}$=7430\,K) of 1.65\,M$_{\sun}$,
was computed using the program developed by Li \& Stix (1994).
Finally, the computed frequencies were compared to the observed ones and
we are led to the results: two theoretical frequencies at
108.25 and 122.22\,$\mu$Hz (i.e. 9.353 and 10.560 d$^{-1}$), corresponding to
$l$=2 and $l$=3 nonradial mode respectively, roughly match
the two observed frequencies (107.98 and 125.18\,$\mu$Hz).
The theoretical pulsation constants for these two model frequencies are
0.050 and 0.045\,d, perfectly agreeing with the observed ones.
This model suggests following physical parameters:
{\em R}=1.95\,R$_{\sun}$, age=0.90\,Gyr,
$\log T_{\rm c}$=7.32 (central temperature), $\log P_{\rm c}$=17.24
(central pressure) and $\log D_{\rm c}$=1.92 (central density).
The radius and age conform to those derived  for UV Tri in the preceding section.
Figure~\ref{Fig:freq-model} shows the computed frequencies compared
with the observed ones.
We note that the suspected term $f_3$ has a counterpart
(at $\sim$16.7 d$^{-1}$ with $l$=3) in the model frequencies.
%
%

\section{Discussion and conclusions}
We present the data of mainly Johnson $V$ and Str\"omgren {\em uvby$\beta$}
for the $\delta$~Sct-type variable UV Tri collected at
the Beijing Astronomical Observatory (China) and
the Observatorio de Sierra Nevada (Spain).
The data were carefully analysed for multiperiodicity and used to
derive stellar physical parameters.
The two frequencies $f_1$=9.3298 d$^{-1}$ and $f_2$=10.8513 d$^{-1}$ were
detected over a commonly acceptable confidence level, confirming the
previous detection.
Besides the two dominating frequencies, some other frequencies higher
than 8 d$^{-1}$
as well as noise peaks in 0--7 d$^{-1}$ domain are visible.
Lower-frequency peaks might be caused by incomplete extinction correction
between variable and comparison stars and from instrumental shifts.
We have no information on the comparison stars' colours and spectral types.
Colour extinction because of different spectral types between the variable
and comparison may also be a cause.
Another possible reason causing the red noise might be the presence of
small amplitude variations of the two frequencies $f_1$ and $f_2$.
Therefore, we gave up period-searching in the lower-frequency region
and only considered higher frequencies.

However, the observed light curves could not be well reproduced by
the synthetic curves with the two frequencies.
There are deviations on some nights. Some data points are undulatory.
These mean that additional pulsation frequencies might present in
the light curves of the variable.
The frequency $f_3$=16.6863 d$^{-1}$ is suspected to be intrinsic and
is also predicted by the theoretic pulsation model, but confirmation is required
because its amplitude is below the confidence level of current detection.

On the other hand, we noted that the poor fit may be
resulted from instrumental, atmospheric and other unknown effects.
We pointed out the phenomenon of anomalous cycles observed in the light curves
of UV Tri similar to that reported for 57 Tau by Papar\'{o} et al.\ (2000) and
further scrutinized by Handler (2001).
We explored the data for possible reasons.
The fact is clear that the two frequencies do not fit these special shapes.
The anomalous cycles seemed not to be caused by non-photometric conditions,
unstable instruments or the comparison star's variability.
As argued by Handler (2001), intrinsic multiperiodicity or
atmospheric/instrumental effects could not explain such a phenomenon.
They seem intrinsic but non-periodic.
In order to examine the phenomenon and derive any regularity,
further observations, especially coordinated multisite observations would be helpful.
The use of different instruments will be a good tool
for excluding the possibility of an instrumental cause.

We further searched the three-year data for possible frequency and
amplitude variability.
The pulsation frequencies were constant within the error bars (see
Table~\ref{Tab:freq-var}). Table~\ref{Tab:ampl-var} shows the amplitudes
in the three years.
We think the presence of anomalous cycles had little effect on
the amplitudes in 2000 and 2001, while no anomalous cycles were observed in 1999.
Within the error ranges, the amplitudes of $f_1$ were constant in 1999--2001.
Concerning the shorter data set in 1999, the corresponding amplitudes
might be suffered from higher but underestimated errors.
However, the amplitude of $f_2$ seemed to increase from 1999 to 2000.

Based on the photometric indices, we derived the physical parameters for UV Tri.
UV Tri is a normal Pop.~I $\delta$ Sct star with solar abundances
evolving on its MS stage before the turn-off point (age=0.89$\pm$0.05~Gyr).
The derived photometric properties and stellar parameters are given in
Table~\ref{Tab:phy-par}.
From these parameters, we calculated the pulsation constants for the two modes.
We compared UV Tri to the models calculated for several well-studied $\delta$ Sct
stars available in the literature, such as FG Vir, XX Pyx and others,
and to the models for radial fundamental and overtone oscillations.
All the existing models consistently identify both modes to be nonradial.
No trace of a radial mode was found in these models for the observed
frequencies of the variable.
An attempt of theoretic pulsation modelling made by us refer
$f_1$ and $f_2$ to be two nonradial modes with $l$=2 and 3, respectively.
The most interesting result is the identification from the models of Fitch (1981),
which advises both frequencies to be $g_1$ modes.
According to the star's location in the colour-magnitude diagram (see
fig.~8 of Rodr\'{\i}guez \& Breger 2001 or fig.~1 of Handler 1999),
UV Tri is located in the overlapped portion of the instability regions of
$\delta$ Sct and $\gamma$ Dor variables.
We recall the spectral type A3, used currently for UV Tri, is in fact
for its nearby star V Tri (Shaw et al. (1983).
Liu et al.\ (2001) inferred A8 or F0 in terms of a relation of
effective temperature, spectral type and absolute magnitude.
According to the Morgan-Keenan (MK) types in the [$m_1$]-[$c_1$] diagram
of Str\"{o}mgren (1966) or the effective temperature-spectral type
scales of Fitzpatrick \& Garmany (1990),
we can refer UV Tri to a type of A8V or F0V.
The spectral type helps us further position the star in the colour-magnitude diagram.
The possible presence of low-degree gravity mode pulsation tends to
suggest UV Tri to be
a $\delta$ Sct star pulsating in nonradial modes similar to $\gamma$ Dor variables.
The possible connection between the $\delta$ Sct and $\gamma$ Dor
variables has been discussed by Breger \& Beichbuchner (1996).
These kind of stars would greatly increase the possibility for
asteroseismology of both classes of variable star.
Furthermore, a $\gamma$ Dor star may pulsate in $\delta$ Sct-like $p$-modes.
A few other stars with surface abundance
peculiarities ($\lambda$ Bo\"{o}tis-type and Am stars) also have been
observed that exhibit $\gamma$ Dor pulsations. These stars with
$g$-mode pulsations offer interesting observational tests for current
$\gamma$ Dor pulsation-driving mechanisms (Guzik et al.\ 2000).

\section*{Acknowledgments}
The authors are very grateful to the anonymous referee for his
constructive comments and many small changes made to the English in this paper,
which helped to improve the manuscript.
This research was funded by the National Natural Science Foundation of
China. Eloy Rodr\'{\i}guez acknowledges partial research support from
the Junta de Andaluc\'{\i}a and the Direccion General
de Investigacion (DGI) under project AYA2000-1559.

\label{lastpage}


\begin{thebibliography}{99}
  \bibitem[\protect\citeauthoryear{Arentoft et al.}{1998}]{aren98}
Arentoft T., Kjeldsen H., Nuspl J., Bedding T. R., Front'{o} A.,
  Viskum M., Frandsen S., Belmonte J. A., 1998, A\&A, 338, 909
  \bibitem[\protect\citeauthoryear{Balona}{1994}]{balo94} Balona L. A.,
1994, MNRAS, 268, 119
  \bibitem[\protect\citeauthoryear{Breger}{1990}]{breg90} Breger M.,
1990, Comm. in Asteroseismology, 20, 1 (Univ. Vienna)
  \bibitem[\protect\citeauthoryear{Breger}{1994}]{breg94} Breger M.,
1994, in Sterken C., de Groot M., eds.,
  NATO ASI Ser. C, Vol. 436, The Impact of Long-term Monitoring on
Variable Star Research:
  Astrophysics, Instrumentation, Data Handling, Archiving. Kluwer,
Dordrecht, p. 393
  \bibitem[2000]{breg00} Breger M., 2000, Baltic Astron., 9, 149
  \bibitem[\protect\citeauthoryear{Breger}{1996}]{breg96} Breger M.,
Beichbuchner F., 1996, A\&A, 313, 851
  \bibitem[\protect\citeauthoryear{Breger}{1993}]{breg93} Breger M. et
al., 1993, A\&A, 281, 90
  \bibitem[\protect\citeauthoryear{Breger}{1999}]{breg99} Breger M.,
Pamyatnykh A. A., Pikall H., Garrido R., 1999, A\&A, 341, 151
  \bibitem[\protect\citeauthoryear{Claret}{1995}]{clar95} Claret A.,
1995, A\&AS, 109, 441
  \bibitem[\protect\citeauthoryear{Crawford}{1979}]{craw79} Crawford D.
L., 1979, AJ, 84, 1858
  \bibitem[\protect\citeauthoryear{Crawford}{1976}]{craw76} Crawford D.
L., Mandwewala N., 1976, PASP, 88, 917
  \bibitem[1975]{deem75} Deeming T. J., 1975, Ap\&SS, 36, 137
  \bibitem[1999]{domi99} Domingo A., Figueras F., 1999, A\&A, 343, 446
  \bibitem[\protect\citeauthoryear{Fitch}{1981}]{fitc81} Fitch W. S.,
1981, ApJ, 249, 218
  \bibitem[2000]{garr00} Garrido R., 2000, in Breger M., Montgomery M.
H., eds.,
  ASP Conf. Ser. Vol. 210, Delta Scuti and Related Stars. Astron. Soc.
Pac.,
  San Francisco, p. 67
  \bibitem[2000]{guzik00} Guzik J. A., Kaye A. B., Bradley P. A., Cox
A. N.,
  Neuforge C., 2000, ApJ, 542, L57
  \bibitem[1999]{hand99} Handler G., 1999, MNRAS, 309, L19
  \bibitem[2001]{hand01} Handler G., 2001, Comm. in Asteroseismology,
140, 6 (Austrian Acad. Sci.)
  \bibitem[1991]{hao99}  Hao J.-X., 1991, Publ. Beijing Astron. Obs.,
18, 35
  \bibitem[1998]{hauc98} Hauck B., Mermilliod M., 1998, A\&AS, 129, 431
  \bibitem[1998]{jian98} Jiang X.-J.,  Hu J.-Y., 1998, Acta Astron.
Sinica, 39, 438
  \bibitem[1994]{li94}  Li Y., Stix M., 1994, A\&A, 286, 811
  \bibitem[1995]{liu95} Liu Z.-L., 1995, A\&AS, 113, 477
  \bibitem[1999]{liu99} Liu Z.-L., Zhou A.-Y., Jiang S.-Y., Liu Y.-Y.
and Li Z.-P.,
    1999, A\&AS, 137, 445
  \bibitem[2000]{liu00} Liu Z. L., Zhou A. Y., Xu D. W., Lu Y.,
  2000, IBVS, No.\,4981
  \bibitem[\protect\citeauthoryear{Liu,Zhou,Rodr\'{\i}guez}{Liu et
al.}{2001}]{liu01} Liu Z.-L., Zhou A.-Y., Rodr\'{\i}guez E., 2001,
  Comm. in Asteroseismology, 140, 52 (Austrian Acad. Sci.)
  \bibitem[1990]{lope90} L\'{o}pez de Coca P., Rolland A.,
Rodr\'{\i}guez E.,
  Garrido R., 1990, A\&AS, 83, 51
  \bibitem[1986]{mala86} Malagnini M. L., Morossi C., Rossi L., Kurucz
R. L. ,
  1986, A\&A, 162, 140
  \bibitem[1999]{mich99} Michel E., Hern\'{a}ndez M. M., Houdek G.,
Goupil M. J., Lebreton Y.,
P\'{e}rez Hern\'{a}ndez F., Baglin A., Belmonte J. A., Soufi F., 1999,
A\&A, 342, 153  
  \bibitem[1999]{montgomery} Montgomery M. H., O'Donoghue D., 1999,
  Delta Sct Star Newsletter, 13, 28 (Univ. Vienna)
  \bibitem[1985]{moon} Moon T. T., Dworetsky M. M., 1985, MNRAS, 217,
305
  \bibitem[1990]{nather90} Nather R. E., Winget D. E., Clemens J. C.,
Hansen J. C.,
    Hine B. P., 1990, ApJ, 361, 309
  \bibitem[1998]{pamy98} Pamyatnykh A. A., Dziembowski W. A., Handler
G., Pikall H., 1998, A\&A, 333, 141  
  \bibitem[2000]{papa00} Papar\'{o} M., Rodr\'{\i}guez E., McNamara B.
J., Koll\'{a}th Z.,
  Rolland A., Gonzalez-Bedolla S. F., S.-Y. Jiang, Z.-P. Li., 2000,
A\&AS, 142, 1
  \bibitem[1972]{pete72} Petersen J. O., J$\o$rgensen H. E., 1972,
A\&A, 17, 367
  \bibitem[1997]{riba97} Ribas I., Jordi C., Torra J., Gim\'{e}nez A.,
  1997, A\&A, 327, 207
  \bibitem[2001]{rodr01} Rodr\'{\i}guez E., Breger M., 2001, A\&A, 366,
178
  \bibitem[1997]{rodr97} Rodr\'{\i}guez E., Gonz\'{a}lez-Bedolla S. F.,
Rolland A., Costa
V., L\'{o}pez-Gonz\'{a}lez M. J., 1997, A\&A, 324, 959
  \bibitem[2000]{rodr00} Rodr\'{\i}guez E., L\'{o}pez-Gonz\'{a}lez M.
J., L\'{o}pez de Coca P.,
 2000, A\&AS 144, 469.
  \bibitem[1982]{scar82} Scargle J., 1982, ApJ, 263, 835
  \bibitem[1983]{shaw} Shaw J. S., Fraquelli D. A., Martins D. H.,
              Stooksbury D. E., 1983, IBVS, No.\,2289
  \bibitem[1993]{smal93} Smalley B., 1993, A\&A, 274, 391
  \bibitem[1997]{smal97} Smalley B., Kupka F., 1997, A\&A, 328, 349
  \bibitem[1998]{sper98} Sperl M., 1998, Comm. in Asteroseismology,
111, 1 (Univ. Vienna)
  \bibitem[1979]{stel79} Stellingwerf R. F., 1979, ApJ, 227, 935
  \bibitem[1997]{temp97} Templeton M. B., McNamara B. J., Guzik J. A.,
Bradley P. A.,
  Cox A. N., Middleditch J., 1997, AJ, 114, 1592      
itself
  \bibitem[2000]{temp00} Templeton M. R., Bradley P. A., Guzik J. A.,
2000, ApJ, 528, 979
  \bibitem[2001]{temp01} Templeton M., Basu S., Demarque P., 2001, ApJ,
submitted (astro-ph/0108458)
  \bibitem[1998]{visk98} Viskum M., Kjeldsen H., Bedding T. R., Dall T.
H.,
  Baldry I. K., Bruntt H., Frandsen S., 1998, A\&A, 335, 549
  \bibitem[1990]{wei90} Wei M.-Z., Chen J.-S., Jiang Z.-J., 1990, PASP,
102, 698
  \bibitem[2001]{zay01}  Zhou A.-Y., 2001, A\&A, 374, 235
  \bibitem[2001]{zhou01} Zhou A.-Y., Rodr\'{\i}guez E., Liu Z.-L.,
   Du B.-T. 2001, MNRAS, 326, 317

\end{thebibliography}
\end{document}